\documentclass[amsmath,amssymb,superscriptaddress]{revtex4-2}
\usepackage{graphicx}
\usepackage{dcolumn}
\usepackage{bm}

\font\scripti=cmmi7
\font\scriptscripti=cmmi5
\def\sib#1{\setbox0 = \hbox{\scripti #1}
  \kern-.02em\copy0\kern-\wd0
  \kern.04em\box0} 
\def\ssib#1{\setbox0 = \hbox{\scriptscripti #1}
  \kern-.02em\copy0\kern-\wd0
  \kern.04em\box0} 
\font\tenib=cmmib10 
\skewchar\tenib='177 \skewchar\tenib='177 \skewchar\tenib='177
\def\pbold#1{\setbox0 = \hbox{$ #1 $}
  \kern-.022em\copy0\kern-\wd0
  \kern.011em\copy0\kern-\wd0
  \kern.011em\copy0\kern-\wd0
  \kern.011em\copy0\kern-\wd0
  \kern.011em\box0} 

\usepackage{graphicx}
\usepackage{dcolumn}
\usepackage{bm}

\def\lesssim{\ \raise.3ex\hbox{$<$}\kern-0.8em\lower.7ex\hbox{$\sim$}\ }
\def\gesim{\ \raise.3ex\hbox{$>$}\kern-0.8em\lower.7ex\hbox{$\sim$}\ }

\newcommand{\red}[1]{{#1}}

\begin{document}
\title{Density-Induced Hadron-Quark Crossover via the Formation of Cooper Triples}
\author{Hiroyuki Tajima}
\affiliation{Department of Physics, Graduate School of Science, The University of Tokyo, Tokyo 113-0033, Japan}
\author{Shoichiro Tsutsui}
\affiliation{QunaSys Inc., Aqua Hakusan Building 9F, 1-13-7 Hakusan, Bunkyo, Tokyo 113-0001, Japan}
\affiliation{RIKEN iTHEMS, Wako, Saitama 351-0198, Japan}
\author{Takahiro M. Doi}
\affiliation{Research Center for Nuclear Physics (RCNP), Osaka University, 567-0047, Japan}
\affiliation{RIKEN iTHEMS, Wako, Saitama 351-0198, Japan}
\author{Kei Iida}
\affiliation{Department of Mathematics and Physics, Kochi University, Kochi 780-8520, Japan}

\date{\today}
\begin{abstract}
We discuss the hadron--quark crossover accompanied by the formation of Cooper triples (three-body counterpart of Cooper pairs) by analogy with the Bose--Einstein condensate to
Bardeen--Cooper--Schrieffer crossover in two-component fermionic systems. Such a crossover is different from a phase transition, which often involves symmetry breaking. We calculate the in-medium three-body energy from the three-body $T$-matrix with a phenomenological three-body force characterizing a bound hadronic state in vacuum. 
With increasing density, the hadronic bound-state pole smoothly undergoes a crossover toward the Cooper triple phase where the in-medium three-body clusters coexist with the quark Fermi sea. 
The relation to the quarkyonic matter model can also be found in a natural manner.
\end{abstract}
\maketitle

\section{Introduction}
After the discovery of massive neutron stars and dramatic development of astrophysical observatories,
the properties of dense matter have attracted great attention~\cite{baym2018hadrons}.
In particular, the equation of state for extremely dense matter has been determined from the astrophysical observations such as $X$-ray and gravitational wave measurements~\cite{RevModPhys.88.021001,baiotti2019gravitational}, and is regarded as an important testing platform of many-body nuclear theories~\cite{lattimer2012nuclear}. 

Along this direction, it may well be an exciting and challenging problem to clarify how nuclear matter changes into possible quark matter in the central core region of massive compact stars.
Such a high-density regime is out of the range accessible by lattice quantum chromodynamics (QCD) simulation due to the severe sign problem~\cite{nagata2022finite}, and there are various predictions such as first-order phase transition and the appearance of quarkyonic matter~\red{\cite{MCLERRAN200783,fukushima2010phase}}.
A fascinating idea to resolve this problem is a crossover scenario called ``hadron--quark crossover''~\red{\cite{masuda2013hadron,10.1093/ptep/ptt045}},
where hadrons consisting of three quarks, i.e., baryons, gradually change into deconfined quarks without phase transitions along the evolution of the chemical potential (or the baryon density).
Interestingly, it is reported that such a crossover can be probed via future observations of the gravitational waves emitted from binary neutron star mergers~\red{\cite{PhysRevLett.129.181101,PhysRevD.106.103027}}.
In particular, the nuclear equation of state constrained from the recent simultaneous measurements of neutron star masses and radii implies a non-monotonic behavior of the speed of sound in the high-density regime~\cite{Kojo_2022}.

Recently, the existence of the peak in the density dependence of the speed of sound has been discussed in the context of hadron--quark crossover~\cite{kojo2021qcd} as well as quarkyonic matter~\cite{PhysRevLett.122.122701}.
However, the elucidation of the microscopic crossover mechanism leading to the peak is still challenging because of the sign problem in the finite-density lattice QCD~\cite{nagata2022finite}.
In this regard,
the peak of the speed of sound in two-color dense QCD, for which sign problems are avoidable even at finite density in the lattice Monte Carlo simulations, has been discussed by analogy with the  Bose--Einstein condensate (BEC) to Bardeen--Cooper--Schrieffer (BCS) crossover~\cite{PhysRevD.105.076001}.
Moreover, the latest results of the lattice simulations also indicate the existence of the peak of the speed of sound in the crossover regime~\cite{iida2020two,10.1093/ptep/ptac137}.
A similar strong-coupling crossover phenomenon has been extensively studied in cold atomic systems~\cite{chen2005bcs,zwerger2011bcs,randeria2014crossover,STRINATI20181,ohashi2020bcs},
where the molecular BEC continuously changes to the BCS-type Cooper pair condensate \red{in a manner accompanied by strong pairing fluctuations~\cite{PhysRevLett.125.060403,universe6110208}} in two-component Fermi gases.
The possible occurrence of the peak of the speed of sound has also been discussed in the density-induced BEC--BCS crossover with the finite-range interaction~\cite{PhysRevA.106.043308}.
We note that the carrier-density-induced BEC--BCS crossover has been realized in strongly correlated electron systems recently~\cite{Kasahara2014PNAS,hashimoto2020bose,nakagawa2021gate,PhysRevX.12.011016}. 

Going back to the three-color QCD at finite density,
we have to consider in-medium three-body correlations unavoidably and seriously because of the formation of hadrons consisting of three quarks~\cite{pittel1990nucleus}, just like the case of the BEC--BCS crossover in which one has to consider in-medium two-body correlations.
Such a three-quark bound state accompanies famous confinement effects associated with the SU(3) gauge invariance~\cite{yagi2005quark,greiner2007quantum}. 
While the three-body correlations in finite-density fermionic systems are still elusive, several ideas appear in the context of cold atomic physics as a quantum simulation of matter in extreme conditions.
Indeed,
although there are several difficulties, such as atomic losses,
three-component ultracold Fermi gases (mimicking three-color quarks) can be prepared experimentally using the three different hyperfine states~\cite{O'Hara_2011}.
Moreover, a similarity between the three-body crossover in three-component Fermi gases and the hadron--quark continuity in the three-color QCD has been pointed out theoretically~\cite{PhysRevLett.109.240401,PhysRevLett.114.115302}.
To understand a non-trivial state of matter in the Fermi-degenerate state with three-body correlations,
one may borrow the knowledge of the BCS--BEC crossover physics well established in cold-atom communities.
In this regard, just as Cooper pairs are formed in the BCS side of the density-induced BEC--BCS crossover~\cite{PhysRevB.60.12410},
one may imagine its three-body counterpart, that is,
a Cooper triple~\cite{PhysRevA.86.013628,PhysRevA.96.053614,PhysRevA.104.053328,PhysRevResearch.4.L012021,PhysRevA.106.043310,guo2022competition}, in a dense Fermi-degenerate regime but with non-negligible three-body correlations.

\begin{figure}[t]
    
    \includegraphics[width=12cm]{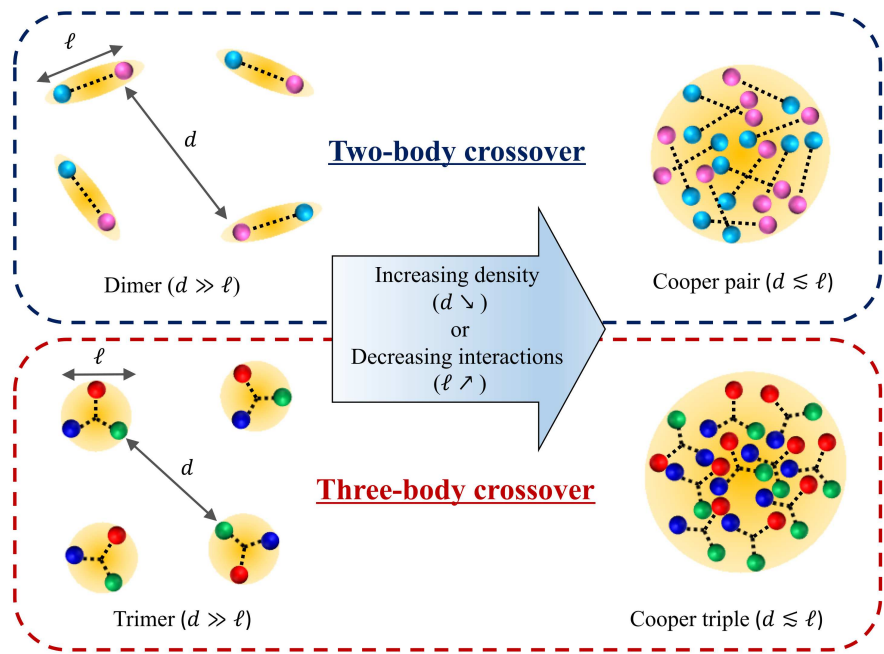}
    \caption{Schematics of the density-induced two- and three-body crossovers.
    While the gases of tightly bound states (i.e., dimers and trimers) exhibit smaller cluster size $\ell$ compared to the mean interparticle distance $d\sim k_{\rm F}^{-1}$ in the low-density regime,
    these bound states start to overlap with each other ($d\lesssim \ell$) with increasing density (i.e., decreasing $d$), and gradually change into Cooper pairs and triples. 
    In ultracold Fermi gases near the Feshbach resonance,
    such an in-medium two-body crossover is triggered by the change in the attractive interaction, where $\ell$ increases when the attractive interaction decreases.
    }
    \label{fig:1}
\end{figure}

Figure~\ref{fig:1} shows a comparison between the two-body crossover (i.e., BEC--BCS crossover) and the three-body crossover accompanied by Cooper triples when the number density increases or the attractive interaction decreases.
First, let us review the famous BEC--BCS-crossover case; if the density increases, the mean interparticle distance given by $d\sim k_{\rm F}^{-1}$ decreases, where $k_{\rm F}$ is the Fermi momentum.
In the dilute regime, $d$ is much longer than the cluster size $\ell$, which is determined by the interaction (e.g., scattering length)~\cite{ohashi2020bcs}.
Such a state is well described by the gas of two-body molecules, and undergoes the BEC at low temperature.
At sufficiently high density, $d$ becomes smaller than the cluster size $\ell$, indicating the tremendous overlap among clusters, that is, the formation of Cooper pairs.
These Cooper pairs are repeatedly formed and dissociated by coexisting with the underlying Fermi sphere, although we do not draw such a  dynamical picture explicitly in Figure~\ref{fig:1}.
While Cooper pairs form the BCS condensate, which spontaneously breaks U(1) gauge symmetry, in conventional superconductors, we note that the spontaneous symmetry breaking is not necessary to form Cooper pairs themselves, as the so-called preformed Cooper pairs are found in strongly correlated or disordered superconductors above the critical temperature~\cite{PhysRevLett.125.097003,bastiaans2021direct}.

Let us now move to the three-body case, as shown in the lower side of Figure~\ref{fig:1}.
One would see tightly bound three-body states with the cluster size $\ell$ if the strong two- or three-body attractive interactions exist in three-color fermionic systems. 
When the number density increases, these trimer states start to overlap at certain densities, and eventually may form an unconventional state that can be analogous to the Cooper pairing state.
In the case of QCD, we recall again that such a three-body state should be kept due to the SU(3) gauge invariance. 
These overlapped three-body states in medium are similar to Cooper triples, which are the three-body counterpart of Cooper pairs predicted in the context of cold atomic physics~\cite{PhysRevA.86.013628,PhysRevA.96.053614,PhysRevA.104.053328,PhysRevResearch.4.L012021,guo2022competition}. 
The Cooper triple state does not necessarily form a condensate (however, note Ref.~\cite{PhysRevA.104.L041302}), but resulting three-body correlations can modify the equation of state substantially~\cite{PhysRevA.102.023313,PhysRevResearch.4.L012021}.

In this paper, motivated by the recent astronomical observations of compact stars as well as the theoretical progress in the in-medium three-body correlations in cold atomic systems,
we discuss the possible scenario of the density-induced hadron--quark crossover accompanied by the formation of Cooper triples using a simplified toy model with three-quark attractive interaction responsible for the baryon formation.
We show that the in-medium three-body energy obtained from the three-body $T$-matrix exhibits a unique crossover from baryonic tightly bound states to Cooper triples with increasing quark chemical potential.

This paper is organized as follows.
In Section~\ref{sec:2},
we introduce the theoretical model and explain how to address the crossover via the three-body $T$-matrix approach~\cite{PhysRevResearch.4.L012021}.
In Section~\ref{sec:3}, we show the numerical result for the in-medium three-body energy and discuss the density-induced hadron--quark crossover.
We summarize this paper in Section~\ref{sec:4}.
Throughout this paper,
we use units in which $\hbar=k_{\rm B}=1$ and the system volume is set to unity.

\section{Model and Methods}
\label{sec:2}
In a way similar to the study of quark--gluon plasmas in Ref.~\cite{PhysRevC.97.034918},
we start from the Hamiltonian based on the Thompson scheme~\cite{PhysRevD.1.110}, as given by
\begin{align}
\label{eq:1}
    H=\sum_{\bm{p}}\sum_{j}\varepsilon_{\bm{p}}\psi_{\bm{p},j}^\dag \psi_{\bm{p},j}
    +\sum_{\bm{k},\bm{q},\bm{k}',\bm{q}',\bm{P}}
    V_{\bm{k},\bm{q},\bm{k}',\bm{q}',\bm{P}}
    \psi_{\bm{k},{\rm r}}^\dag
    \psi_{\bm{q},{\rm g}}^\dag
    \psi_{\bm{P}-\bm{k}-\bm{q},{\rm b}}^\dag
    \psi_{\bm{P}-\bm{k}'-\bm{q}',{\rm b}}
    \psi_{\bm{q}',{\rm g}}
    \psi_{\bm{k}',{\rm r}},
\end{align}
where $\psi_{\bm{p},j}^{(\dag)}$ is the fermionic annihilation (creation) operator with three colors ($j={\rm r,g,b}$), and
$\varepsilon_{\bm{p}}$ is the relativistic dispersion $\varepsilon_{\bm{p}}=\sqrt{p^2+m_q^2}$ with
the nonzero constituent quark mass $m_q\simeq 0.34$ GeV.
For simplicity, we neglect the spin and flavor degrees of freedom, which are not important for our purpose and will be considered in the future work.
The second term of Equation~(\ref{eq:1}) is the three-body attractive interaction responsible for the color-singlet baryon formation.
While the coupling constant should be momentum-dependent due to the color confinement~\cite{PhysRevD.34.2809,PhysRevLett.86.18,PhysRevD.65.114509} as well as relativistic effects (the projection operator for the positive energy states and the coupling with negative-energy states)~\cite{PhysRevD.1.110,PhysRevC.97.034918}, 
we employ the constant coupling $V_{\bm{k},\bm{q},\bm{k}',\bm{q}',\bm{P}}\simeq V_3$ for the purpose of qualitatively understanding the hadron--quark crossover accompanied by the Cooper triple formation in the relatively high-density regime.
We will adjust the value of $V_{3}$ later in such a way as to roughly reproduce the baryon mass in vacuum $M_{\rm B}$ as $\simeq 0.91$ GeV.
Incidentally, we note that non-relativistic or relativistic quark models with three-body forces have been introduced in Refs.~\cite{RICHARD1983267,PhysRevD.27.233,blask1990hadron}.
We obtain the in-medium three-body $T$-matrix at the zero center-of-mass momentum~\cite{PhysRevResearch.4.L012021},
\begin{align}
    T_3(\Omega)=\frac{V_3}{1-V_3\Xi(\Omega)},
\end{align}
where $\Omega_+=\Omega+i\delta$ is the three-body energy with a small imaginary part $i\delta$.
The in-medium three-body propagator $\Xi(\Omega)$ reads~\red{\cite{PhysRevC.81.064310,BEYER200133}}
\begin{align}
    \Xi(\Omega)&=\sum_{\bm{k},\bm{q}}
    \frac{(1-f_{\bm{k}})(1-f_{\bm{q}})(1-f_{\bm{k}+\bm{q}})+f_{\bm{k}}f_{\bm{q}}f_{\bm{k}+\bm{q}}}{\Omega_++3\mu-\varepsilon_{\bm{k}}-\varepsilon_{\bm{q}}-\varepsilon_{\bm{k}+\bm{q}}},
\end{align}
where
\begin{align}
    f_{\bm{k}}=\frac{1}{\exp\left(\frac{\varepsilon_{\bm{k}}-\mu}{T}\right)+1}
\end{align}
is the Fermi distribution function.
We note that while the antiquarks are ignored in our toy model, such antiparticle contributions involve the antiparticle distribution function $\tilde{f}_{\bm{k}}=\frac{1}{\exp\left(\frac{\varepsilon_{\bm{k}}+\mu}{T}\right)+1}$, which can be small at large chemical potential and low temperature.

We set $V_3=-2.16\times 10^2$ $({\rm GeV})^{-5}$ such that 
$T_3(\Omega)$ has a pole at $\Omega\simeq M_{\rm B}$ in vacuum.
The cutoff and the small imaginary energy for the analytic continuation are adapted as $\Lambda=2$ GeV and $\delta=2\times 10^{-2}$ GeV.
We note that the continuum bottom of three quarks is given by
\begin{align}
    E_{\rm th}=3m_q-3\mu.
\end{align}
Then, the in-medium three-body binding energy $E_{\rm B}^{\rm M}$ (i.e., energy gain due to the formation of the in-medium three-body cluster) can be defined as
\begin{align}
    \frac{1}{V_3}-\Xi(\Omega_{\rm pole}=E_{\rm th}-E_{\rm B}^{\rm M})=0,
\end{align}
which is the pole of the in-medium three-body $T$-matrix.
In the vacuum case, such an energy gain is given by $E_{\rm B}=3m_q-M_{\rm B}\simeq 0.11$ GeV.

\section{Results and Discussion}
\label{sec:3}

\begin{figure}[t]
    
    \includegraphics[width=8cm]{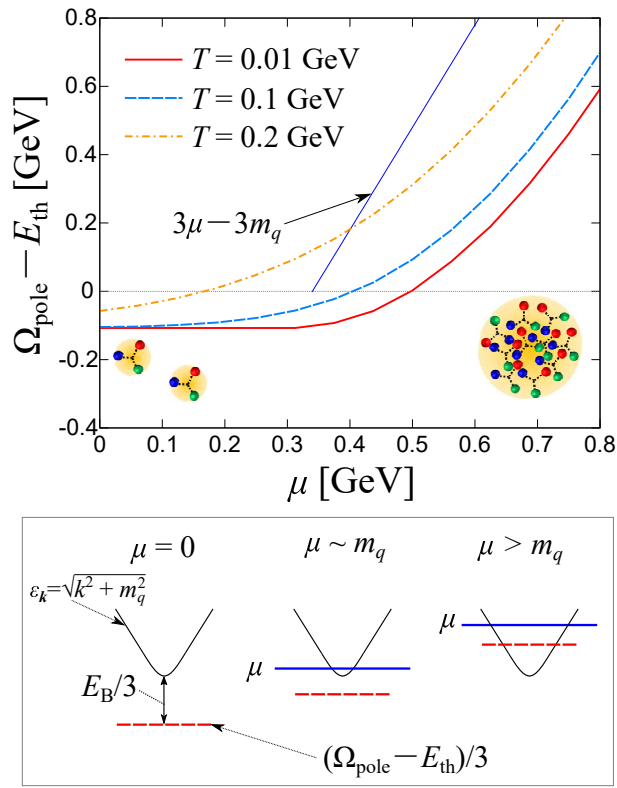}
    \caption{The in-medium 
  three-body energy 
    in the hadron--quark crossover regime
    at different temperatures. For reference, we plot $\Omega-E_{\rm th}=3\mu-3m_q$, that is, $\Omega=0$ (at the Fermi level).
    The lower panel presents the schematic comparison between the single-particle energy $\varepsilon_{\bm{k}}=\sqrt{k^2+m_q^2}$ and the three-body energy per particle $(\Omega_{\rm pole}-E_{\rm th})/3$ (dashed line) along the hadron--quark crossover.}
    \label{fig:2}
\end{figure}

Figure~\ref{fig:2} shows the in-medium three-body energy pole $\Omega_{\rm pole}$ measured from the bottom of the continuum $E_{\rm th}=3m_q-3\mu$ as a function of $\mu$ at different temperatures $T$.
One can see that the the three-body energy located in $\Omega_{\rm pole}-E_{\rm th}\simeq M_{\rm B}-3m_q\equiv -E_{\rm B}$ (i.e., $E_{\rm B}^{\rm M}\simeq E_{\rm B}$) at $\mu=0$ moves toward the continuum bottom $\Omega_{\rm pole}-E_{\rm th}=0$ when $\mu$ increases.
Moreover, this three-body pole remains at larger $\mu$ and locates below the Fermi level given by $\Omega-E_{\rm th}=3\mu-3m_q$, indicating the coexistence of the quark Fermi sea and in-medium three-body clusters.
This result manifests a picture of the smooth crossover from the tightly bound baryonic trimer state to the Cooper triple state, 
being analogous to the density-induced BEC--BCS crossover~\cite{PhysRevB.60.12410,PhysRevC.82.024911,STRINATI20181}.
While the existence of the positive pole in the continuum
is in contrast to the non-relativistic one-dimensional case~\cite{PhysRevResearch.4.L012021},
the similar behavior of the three-body energy can be found in the variational approach to the non-relativistic Fermi gases in three dimensions~\cite{PhysRevA.104.053328}.
At higher temperatures, $\Omega_{\rm pole}$ is shifted toward an upper energy level around the crossover because of the thermal agitation effect on the binding energy~\cite{PhysRevResearch.4.L012021,Tajima_2019}.
However, one can see that the three-body pole is still located below the Fermi level ($\Omega-E_{\rm th}<-E_{\rm th}$, i.e., $\Omega<0$) at low temperatures, implying the presence of the stable three-body bound state in medium.

To see the physical picture of this crossover from a different point of view,
we show the schematic single-particle energy level diagram in the lower panel of Figure~\ref{fig:2}.
At $\mu=0$, the three-body bound state is located in the energy level $-E_{\rm B}/3$ measured from the continuum bottom $\varepsilon_{\bm{k}=\bm{0}}=m_q$.
At finite $\mu$, this three-body bound state is shifted toward an upper energy level because of the reduced binding energy associated with the Pauli blocking effect.
However, such a Fermi surface effect does not suppress the three-body pole, but eventually leads to the in-medium bound state in continuum at high density, as in the case of the famous Cooper pairing in superconductors. This state can be regarded as the Cooper triple phase, but we note that there are no clear boundaries between the hadronic bound-trimer phase and Cooper triple phase, a feature similar to the BEC--BCS crossover~\cite{PhysRevA.104.053328,PhysRevResearch.4.L012021}.

\begin{figure}[t]
    \includegraphics[width=8cm]{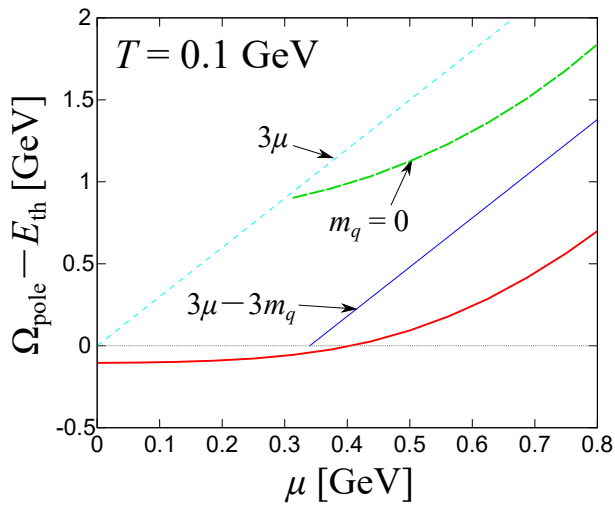}
    \caption{ The in-medium three-body energy with  $m_q=0.34$ GeV and $m_q=0$.
    }
    \label{fig:3}
\end{figure}

While we have fixed $m_q=0.34$ GeV in Figure~\ref{fig:2}, it may exhibit density dependence through the quark--antiquark pairing~\cite{PhysRevLett.120.222001}.
In Figure~\ref{fig:3},
we show a comparison between the results with massive ($m_q=0.34$ GeV) and massless ($m_q=0$) quarks; in the latter case, the chiral symmetry is assumed to be restored due to the Pauli blocking effect on the quark--antiquark pairing.
While the result with $m_q=0$ (dashed curve) is found to be larger than that with $m_q=0.34$ GeV (solid curve), this difference can be approximately given by $3m_q$ with $m_q=0.34$ GeV. 
On the other hand, even the massless case is well below the  corresponding Fermi level ($3\mu$ in the absence of $m_q$) at high density, where the Cooper triple phase is expected to occur.
For more realistic case, $m_q$ changes from the in-vacuum value down to the current quark mass across the evolution of $\mu$, leading to the in-medium three-body pole in between.

In the high-density regime, the three-body pole measured from the Fermi level (i.e., $3\mu-3m_q-(\Omega_{\rm pole}-E_{\rm th})\equiv-\Omega_{\rm pole}$) can be regarded as the energy gain of the formation of a Cooper triple. The ratio of
this energy gain to the Fermi energy may well become smaller as the density increases.
However, the non-local properties of the three-body force, if remaining, would be important in such a high-density regime.
Indeed, if the three-body force is replaced by the linear potential $V_{qqq}(R)\simeq \sigma R$~\cite{PhysRevD.34.2809,PhysRevLett.86.18,PhysRevD.65.114509}, where $\sigma$ is the string tension and $R$ is the characteristic length scale for the three-body state,
the energy gain might be corrected due to a significant spatial overlap of Cooper triples.
Combined effects of the density dependence of the dynamical quark mass and of the linear-shaped three-quark interaction would cause the three-body energy pole to increase more rapidly.
This will be left for an interesting future work.
We remark in passing that at asymptotically high density, the confining three-body force would be suppressed, while two-body interactions due to one-gluon exchange would take over. Even in the density regime of interest here, instanton-induced two-body interactions would take effect.  Such two-body interactions would lead to quark Cooper pairs and hence color superconductivity~\cite{RevModPhys.80.1455}, which are ignored in the present analysis.

\begin{figure}[t]
    \includegraphics[width=12cm]{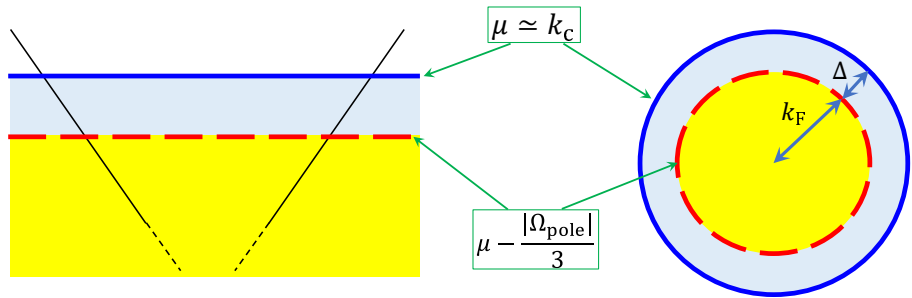}
    \caption{Correspondence between our model and quarkyonic matter proposed in Ref.~\cite{PhysRevLett.122.122701}.
    While the left figure shows the quark dispersion (solid line) with the threshold of Cooper triples $\mu-|\Omega_{\rm pole}|/3$,
    the right figure shows the momentum configuration of quarkyonic matter, where the baryonic excitation (analogous to Cooper triples) can be found in the Fermi shell of baryons $\Delta$ near the quark Fermi surface.
    }
    \label{fig:4}
\end{figure}

\red{
Finally, to see the relevance of our results to neutron star physics,
we discuss the connection between the present work and the phenomenological model of quarkyonic matter developed by McLerran and Reddy~\cite{PhysRevLett.122.122701}, where the peak structure of the speed of sound has been demonstrated as expected from recent astrophysical observations~\cite{Kojo_2022}.
In the high-density regime, the baryonic excitation near the quark Fermi surface can be regarded as the Cooper triples.
Indeed, in this model, the quark number density may be given by
\begin{align}
\label{eq:MR}
    \rho=\frac{g_i}{6\pi^2}\left[k_{\rm F}^3+k_{\rm c}^3-(k_{\rm c}-\Delta)^3\right],
\end{align}
where $g_i$ and $\Delta$ are the degeneracy (e.g., spin and isospin degrees of freedom) and the energy shell of baryons (corresponding to the Debye-frequency-like window of Cooper triples~\cite{PhysRevA.104.L041302}), respectively.
$k_{\rm c}=k_{\rm F}+\Delta\simeq \mu$ is the upper bound of the momentum for the formation of Cooper triples.
While $k_{\rm c}$ is called the Fermi momentum of baryons in Ref.~\cite{PhysRevLett.122.122701},
$k_{\rm c}$ is not necessarily equal to such a momentum in our model because Cooper triples are no longer point-like fermions in the high-density regime.
However, $k_{\rm c}$ plays a similar role in the two different models. 
The last two terms in the parenthesis of Equation~\eqref{eq:MR}, which are regarded as the quark number density in baryons in Ref.~\cite{PhysRevLett.122.122701}, correspond to the quark number density in Cooper triples. 
Let us now set $m_q$ to zero for simplicity.
Then, the Fermi degenerate state can be filled up to the threshold of Cooper triples given by $\frac{\Omega_{\rm pole}-E_{\rm th}}{3}\equiv\mu+\frac{\Omega_{\rm pole}}{3}$
(noting that $3\mu+\Omega_{\rm pole}<3\mu$, and thus, $\Omega_{\rm pole}<0$ at high density, as shown in Figure~\ref{fig:3}).
Therefore, as depicted in Figure~\ref{fig:4}, one may find analogical relations between the present system and the phenomenological model in Ref.~\cite{PhysRevLett.122.122701} as
\begin{align}
    k_{\rm F}\simeq \mu-\frac{|\Omega_{\rm pole}|}{3},
    \quad
    \Delta\simeq \frac{|\Omega_{\rm pole}|}{3}.
\end{align}}
\red{In this way, these results imply that the possible peak in the speed of sound can be regarded as the consequence of the Cooper triple formation.
Moreover, we also note that a related anomaly of the compressibility can be found near the region where the in-medium three-body binding energy rapidly changes in one-dimensional Fermi gases with the scale anomaly~\cite{PhysRevResearch.4.L012021}.
However, detailed investigations of thermodynamic properties within more realistic models would be needed to confirm such a scenario, which is left for interesting future work.
}








\section{Conclusions}
\label{sec:4}
In this paper, we have discussed the density-induced hadron--quark crossover via the formation of Cooper triples by analogy with BEC--BCS crossover realized in cold atomic and in condensed matter systems.
Applying the in-medium three-body $T$-matrix approach to the toy model, where relativistic quarks interact with each other via the phenomenological three-body forces leading to the baryonic three-body bound state in vacuum,
we show that the in-medium three-body pole exhibits a smooth crossover from the state of baryonic three-body trimers to the state of Cooper triples with increasing chemical potential, which can be regarded as the three-body counterpart of the density-induced BEC--BCS crossover. 
\red{A relation of the present system with the quarkyonic matter model~\cite{PhysRevLett.122.122701} exhibiting the peak of the speed of sound has also been discussed.}

For future perspective,
it is interesting to examine the equation of state and the speed of sound in a more realistic model including, e.g., density dependence of the dynamical quark mass,
\red{Pauli blocking effects among bound clusters~\cite{PhysRevD.34.3499}, intercluster repulsion due to quark exchange~\cite{particles3020033},}
effects of gluons, and competition with diquark formation and color superconductivity~\cite{RevModPhys.80.1455}.
\red{In the presence of diquarks, which play a role in the baryon
structure in vacuum~\cite{BARABANOV2021103835,sym12091468}, and the axial anomaly~\cite{PhysRevLett.97.122001}, the Nambu--Gorkov formalism
, which helps avoid the breakdown of the $T$-matrix approach~\cite{kadanoff2018quantum,WANG2011347}, would be useful.}
In the presence of additional degrees of freedom such as spin and flavor, dibaryonic superconducting quark matter might also appear at low temperature~\cite{BARROIS1977390,PhysRevD.105.103005}.
For ab initio calculations to elucidate the hadron--quark crossover, the complex Langevin method would be one of the promising methods to address finite-density QCD by avoiding the sign problem, as demonstrated in Ref.~\cite{PhysRevD.98.114513,ito2020complex}.
The complex Langevin method can also be used for clarifying multi-body correlations in ultracold Fermi gases~\cite{BERGER20211}.

\acknowledgments
The authors thank E. Nakano for useful discussions.
This research was funded by Grants-in-Aid for Scientific
Research provided by JSPS through No.~18H05406, No.~20K14480, No.~22H01158, and No.~22K13981.

\bibliographystyle{apsrev4-2}
\bibliography{reference.bib}

\end{document}